\documentclass[12pt,fleqn,twoside]{article}
\usepackage{latexsym,amsfonts,amssymb}

\newcommand{\be}[1]{\begin{equation}\label{#1}}
\newcommand{\ee}{\end{equation}}
\newcommand{\bs}[1]{\begin{equation}\label{#1}\arraycolsep=0em\begin{array}{l}}
\newcommand{\es}{\end{array}\end{equation}}
\newcommand{\bss}{\[\arraycolsep=0em\begin{array}{l}}
\newcommand{\ess}{\end{array}\]}
\newcommand{\ba}[1]{\begin{array}{#1}}
\newcommand{\ea}{\end{array}}
\newcommand{\bc}{\begin{center} }
\newcommand{\ec}{\end{center} }

\renewcommand{\le}{\leqslant}
\renewcommand{\ge}{\geqslant}

\newcommand{\p}{\partial}
\newcommand{\R}{{\mathbb R}}
\newcommand{\N}{{\mathbb N}}

\begin{document}

\begin{flushleft}
\bf Roman O. Popovych${}^\dag$ and  Vyacheslav M. Boyko${}^\ddag$
\end{flushleft}

\begin{flushleft}
Institute of Mathematics, National Academy of Science of Ukraine,
\\ 3 Tereshchenkivs'ka Street, 01601, Kyiv-4, Ukraine\\
$\dag$~E-mail: rop@imath.kiev.ua\\
$\ddag$~E-mail: boyko@imath.kiev.ua
\end{flushleft}

\begin{flushleft}
\Large\bf Differential Invariants\\
 and  Application to Riccati-Type Systems\par
\end{flushleft}

\begin{abstract}
We generalize the classical Lie results on a basis
of differential invariants for
a one-parameter group of local transformations to the
case of arbitrary number of independent and dependent  variables.
It is proved that if universal invariant of a one-parameter group
is known then a complete set of functionally independent differential
invariants can be constructed
via one quadrature and differentiations. Some applications of first-order differential
invariants to Riccati-type systems are also presented.
\end{abstract}

\vspace{1ex}


\vspace{1ex}


\newpage

\section{Introduction}
The theory of differential invariants nowadays undergoes active development and
is widely used for integration in quadratures and for
order lowering of ordinary differential equations,
and also for description of classes of invariant differential equations \cite{olverEI&S95,FSS}.
In the theory of differential invariants a major role is played
by various versions of the conjecture on the finite basis of differential invariants
that could be non-rigorously formulated in the following way:
{\it for an arbitrary group $G$ of local transformations there exists
such finite set of differential invariants that every differential invariant
of the group $G$ can be represented as a function of these invariants and their derivatives.}
A statement of such type (for one-parameter group of local transformations
in the space of two variables)
was proved by S.~Lie himself~\cite{lie}
(see also \cite{ibragimov,olver.eng})
and soon afterwards it was essentially generalized by A.~Tresse~\cite{tresse}.
The recent progress in this direction is due to the
 works by L.V.~Ovsyannikov \cite{ovsyannikov.eng} and
P.~Olver~\mbox{\cite{olverEI&S95,olver94,olver95,olver97}}, where the notions of the operator of
invariant differentiation, differential invariant coframe etc. are introduced,
and results on rank
stabilization of the prolonged group action and on the estimates for the number of
differential invariants are obtained.
There is also considerable number of papers devoted to the
search for the differential invariants of specific groups (see e.g.~\cite{dif-inv1,dif-inv2,dif-inv3,dif-inv4,dif-inv5}).

In the present paper we study the differential invariants
for one-parameter group of local transformations
in the space of $n$ independent and $m$ dependent variables ($m,n\in\N$).
(The case $n=m=1$ was considered in~\cite{eisenhart,aksenov.ru}
and in our work \cite{boyko-popovych}. The results for the case $n=1$ with arbitrary $m\in{\mathbb N}$
were published in \cite{samara}.)
The importance of this problem stems from its being a part of the problem of
search for differential invariants of a group of arbitrary dimension~\cite{olverEI&S95,olver.eng}.
Our generalization of the Lie theorem on differential invariants of a one-parameter
group of local transformations was done not only with respect to
the number of independent and dependent variables, but also with respect
to strengthening of its statements. It was proved that if a differential
invariant is known, then a complete set of functionally independent
differential invariants can be constructed through one quadrature
and differentiations. As a side product, we also present some
results on the existence of basis of differential invariants rational in
higher jet coordinates. A link between differential invariants and integration of systems of Riccati-type
equations was analyzed. Let us note that results of this paper
can be generalized for some classes
of multiparameter groups of local transformations (or Lie algebras of differential
operators).

\section{Generalization of Lie theorem\\ on differential invariants}
Let
$Q=\xi^a(x,u)\p_{x_a}+\eta^i(x,u)\p_{u^i}$ be an infinitesimal operator of
a one-parameter group
$G$ of local transformations which act on the set $M\subset J_{(0)}=X\times U$, where
$X\simeq \R^n$ is the space of independent variables $x=(x_1,x_2,\ldots, x_n)$ and
$U\simeq \R^m$ is the space of dependent variables $u=(u^1, u^2, \ldots, u^m)$,
$G^{(r)}$ is a prolongation of the action of the group $G$ for the subset
$M_{(r)}=M\times U^{(1)}\times U^{(2)}\times\cdots \times U^{(r)}$
of the jet space $J_{(r)}=X\times U_{(r)}$ of $r$-th
order jets over the space $X\times U$
(here $U_{(r)}=U\times U^{(1)}\times U^{(2)}\times\cdots \times U^{(r)}$, $r\ge 1$,
$Q^{(r)}$ is the $r$-th prolongation of $Q$
\cite{olverEI&S95,olver.eng}).
A function $I\mbox{:}\: M_{(r)}\to {\mathbb R}$ is called
a differential invariant of the
order $r$ for the group $G$ (or for the operator
$Q$) if it is an invariant of
the prolonged action of $G^{(r)}$ (of $Q^{(r)}$). A necessary and sufficient
condition for the function $I$ to be an $r$-th order differential
invariant of the group $G$ is the equality~$Q^{(r)}I=0$.

Here and below, if not otherwise stated,
the indices $a$, $b$, $c$, $d$ run from 1 to $n$,
indices $i$, $j$, $k$, $l$ run from 1 to $m$.
The summation over the repeated indices is understood.

Let $I=I(x,u)=(I^1(x,u),I^2(x,u), \ldots, I^{m+n-1}(x,u))$ be a complete set
of functionally independent invariants
(or {\it a universal invariant} \cite{ovsyannikov.eng})
for an operator $Q$, and $J(x,u)$ is a particular solution of the equation $QJ=1$.
Then the functions $I^1(x,u)$, $I^2(x,u)$, \dots, $I^{m+n-1}(x,u)$  and $J(x,u)$
are functionally independent.
Let us make a local change of variables:
\mbox{$y_c=I^c(x,u)$}, $c=\overline{1,n-1}$, $y_n=J(x,u)$ are new
independent variables, and
$v^i=I^{i+n-1}(x,u)$ are new dependent variables. In terms of variables
$y=(y_1, y_2, \ldots, y_n)$ and $v=(v^1, v^2, \ldots, v^m)$
the operator $Q$ has the form $\p_{y_n}$.
Thus for any $r\ge 1$ the form of the prolonged operator~$Q^{(r)}$
coincides with $Q=\p_{y_n}$, and therefore
\bss
\hat y=(y_1, y_2, \ldots, y_{n-1}), \\[1ex]\displaystyle\left.
v_{(r)}=\left\{v^i_\alpha\!=\!\frac{\p^{|\alpha|}v^i}
{\p y_1^{\alpha_1}\p y_2^{\alpha_2}\ldots\p y_n^{\alpha_n}}\;\right|\:
\alpha_a\!\in\!\N\!\cup\!\{0\},\;|\alpha|\!=\!\!\sum_{a=1}^n\alpha_a\!\le\! r
\right\}
\ess
(here $v^i_\alpha=v^i$, if $|\alpha|=0$)
form a complete set of its functionally independent
invariants, and $(\hat y, v_{(r)})$~is a universal invariant of the group~$G^{(r)}$.
(Functional independence of the components $\hat y$ and $v_{(r)}$
is obvious, as $(y,v_{(r)})$ is a set of variables in the space~$J_{(r)}$.)
This means that $(\hat y, v)$ is a fundamental set of differential invariants for the operator
$Q$, i.e. any differential invariant of the operator
$Q$ can be represented as a function of $\hat y$ and $v$ and of the derivatives
of $v$ with respect to
operators of $G$-invariant differentiation. These operator coincide here with the operators
$D_{y_a}=\p_{y_a}+v^i_{y_a}\p_{v^i}+v^i_{y_ay_b}\p_{v^i_{y_b}}+\cdots$
of total derivatives with respect to the variables~$y_a$.

Let us go back to the variables $x,$ $u$. In terms of these variables
\bs{Dy.in.xu}\displaystyle
D_{y_c}= \frac{(-1)^{c+a}}{\Delta}
\frac{D(I^d,{\scriptstyle d=\overline{1,n-1}}, {\scriptstyle d\not=c},J)}
{D(x_b, {\scriptstyle b=\overline{1,n}}, {\scriptstyle b\not=a})}D_{x_a},
\quad c=\overline{1,n-1},
\\[2.5ex] \displaystyle
D_{y_n}= \frac{(-1)^{n+a}}{\Delta}
\frac{D(I^d,{\scriptstyle d=\overline{1,n-1}})}
{D(x_b, {\scriptstyle b=\overline{1,n}}, {\scriptstyle b\not=a})}D_{x_a},
\es
where $D_{x_a}=\p_{x_a}+u^i_{x_a}\p_{u^i}+u^i_{x_ax_b}\p_{u^i_{x_b}}+\cdots$~is the
operator of total derivative with respect to the variable~$x_a$, and
\[
\frac{D(I^d,{\scriptstyle d=\overline{1,n-1}}, {\scriptstyle d\not=c}, J)}
{D(x_b, {\scriptstyle b=\overline{1,n}}, {\scriptstyle b\not=a})},
\;
\frac{D(I^d,{\scriptstyle d=\overline{1,n-1}})}
{D(x_b, {\scriptstyle b=\overline{1,n}}, {\scriptstyle b\not=a})},
\;
\Delta=\frac{D(I^d, {\scriptstyle d=\overline{1,n-1}}, J)}
{D(x_b, {\scriptstyle b=\overline{1,n}})}
\]
denote Jacobians (of total derivatives) \\
\hspace*{5mm} of the functions $I^d$, $d=\overline{1,n-1}$, $d\not=c$,
$J$ with respect to the variables~$x_b$, $b=\overline{1,n}$, $b\not=a$,\\
\hspace*{5mm} of the functions $I^d$, $d=\overline{1,n-1}$
with respect to the variables~$x_b$, $b=\overline{1,n}$, $b\not=a$,\\
\hspace*{5mm} of the functions $I^d$, $d=\overline{1,n-1}$, $J$
with respect to the variables~$x_b$, $b=\overline{1,n}$,\\ respectively.

As a result, we arrive at the following theorem.

\medskip

\noindent
{\bf Theorem 1.} {\it Let $I(x,u)=(I^1(x,u),I^2(x,u), \ldots, I^{m+n-1}(x,u))$ be a
universal invariant of an operator $Q$ and
$J(x,u)$ be a particular solution of the equation $QJ=1$. Then functions
\[
I^c(x,u), \quad
D_{y_1}^{\alpha_1}D_{y_2}^{\alpha_2}\ldots D_{y_n}^{\alpha_n}I^{i+n-1}(x,u), \quad
\]
where $c=\overline{1,n-1}$, $\alpha_a\!\in\!\N\cup\{0\},$ $\sum_{a=1}^n\alpha_a\le r$,
and operators $D_{y_a}$ are determined by the formulae~(\ref{Dy.in.xu}),
form a complete set of functionally independent $r$-th order differential
invariants (or a universal differential invariant)
for the operator $Q$.}

\medskip

\noindent
{\bf Corollary 1.} {\it For any operator $Q$ there exists a
complete set of functionally independent $r$-th order differential invariants,
where every invariant is a rational function of the variables
$u^i_\alpha$
($\alpha=(\alpha_1,\alpha_2,\ldots,\alpha_n),$
$\alpha_a\in\N\cup \{0\}$, $0<\sum_{a=1}^n\alpha_a\le r$) of the jet space $J_{(r)}$
with~coefficients depending on  $x_a$ and $u^j$.}

\medskip

\noindent
{\bf Corollary 2.} {\it If $I=(I^1(x,u),I^2(x,u), \ldots, I^{m+n-1}(x,u))$~is a
universal invariant for the operator $Q$  and
$J=J(x,u)$~is a particular solution for the equation $QJ=1$, then the functions
\bs{first.order.n.m.diff.invariants}\displaystyle
D_{y_c}I^{i+n-1}= \frac{(-1)^{c+a}}{\Delta}
\frac{D(I^d,{\scriptstyle d=\overline{1,n-1}}, {\scriptstyle d\not=c},J)}
{D(x_b, {\scriptstyle b=\overline{1,n}}, {\scriptstyle b\not=a})}D_{x_a}I^{i+n-1},
\\[2.5ex] \displaystyle
D_{y_n}I^{i+n-1}= \frac{(-1)^{n+a}}{\Delta}
\frac{D(I^d,{\scriptstyle d=\overline{1,n-1}})}
{D(x_b, {\scriptstyle b=\overline{1,n}}, {\scriptstyle b\not=a})}D_{x_a}I^{i+n-1}
\es
($c=\overline{1,n-1}$) form a complete set of functionally independent
 differential invariants having exactly order one for the operator $Q$.}

\medskip

Note that if a universal invariant~$I$ for the operator~$Q$ is known, then
a particular solution for the equation $QJ=1$ may be easily found
via one quadrature. E.g. if for some fixed
$a$ $\xi^a\not=0$, we have a particular solution
\[\textstyle J(x,u)=\int dx_a/\xi^a(X^1\ldots,X^{a-1},x_a,X^{a+1},\ldots,
X^n,U^1,\ldots,U^m),\]
where $x_b=X^b(x_a,C)$, $b\not=a$, $u^j=U^j(x,C)$~is the solution for
the system of algebraic equations
$I(x,u)=C:=(C_1,C_2,\ldots,C_{m+n-1})$
with respect to the variables $x_b$, $b\not=a$, $u^j$,
and after the integration it is necessary to perform an inverse substitution
$C=I(x,u)$ (there is no summation with respect to $a$ in this case).
Likewise, when $\eta^i\not=0$ for some fixed $i$, then we can assume
\[\textstyle J(x,u)=\int du^i/\eta^i(X^1,\ldots,X^n,U^1,\ldots,U^{i-1},u^i,U^{i+1},
\ldots,U^m)\]
(there is no summation with respect to $i$), where
$x_b=X^b(u^i,C)$, $u^j=U^j(u^i,C)$, $j\not=i$,~is the solution of
the system of algebraic equations
$I(x,u)=C$ with respect to the variables $x_b$, $u^j$, $j\not=i$.

Thus, the following theorem holds.

\medskip

\noindent
{\bf Theorem 2.} {\it If a universal invariant is found for the operator $Q$, then
a complete set of functionally independent differential
invariants of arbitrary order
may be constructed through one quadrature and differentiations.}

\section{Invariant Differentials}

Let us introduce a notion of an invariant differential
that is a particular case of a more general notion
of a first-order contact invariant differential form in the jet space~\cite{olver95}.

\medskip

\noindent
{\bf Definition.} {\it A differential $dW(x,u)$ will be called {\it invariant}
with respect to a group~$G$ (with operator $Q$), if it does not change under action of
transformations from the group~$G$.}

\medskip

A criterion for invariance of a differential is an equality
$dQW(x,u)=0.$ Two essentially different cases are possible:

\vspace{0.5ex}

\noindent
1) the function $W(x,u)$ is an invariant of the operator $Q$, i.e. $QW(x,u)=0$;
than the differential $dW(x,u)$ is automatically invariant with respect to
the operator~$Q$ ({\it invariant differential of the first type});

\vspace{0.5ex}

\noindent
2) the function $W(x,u)$ is not invariant under $Q$,
while the differential
 $dW(x,u)$ is ({\it invariant differential of the second type});
then $QW(x,u)$~is a non-zero constant.

\vspace{0.5ex}

If a set of functions $I(x,u)=(I^q(x,u))_{q=\overline{1,m+n-1}}$ and $J(x,u)$,
that determine a universal invariant of the operator $Q$ and
the invariant  differential of the second type is type, then all such sets may be
found according to the formulae
\be{equivalence.relation.on.invariant.differentials}
\hat I(x,u)=F(I(x,u)),\qquad \hat J(x,u)=J(x,u)+H(I(x,u)),
\ee
where $F=(F^1,F^2,\dots,F^{m+n-1})$ and $H$~are differentiable functions of their arguments,
\mbox{$|\p F/\p I|\not=0$}.
The formulae~(\ref{equivalence.relation.on.invariant.differentials}) determine
the equivalence relation $\Omega$ on the set $\cal M$ of collections of
$m+n$ smooth functions of $m+n$ variables with a non-zero Jacobian.
We denote the corresponding set of equivalence classes as ${\cal M}/\Omega$.

\medskip

\noindent
{\bf Proposition.} {\it There is a one-to-one correspondence between ${\cal M}/\Omega$
 and the set of non-zero operators $\{Q\}$ in the space of variables $(x,u)$: the set
$\{(I(x,u);J(x,u))\}$ of solutions of the system
$QI^q=0$, $q=\overline{1,m+n-1}$, $QJ=1$, where $I^q$~are functionally independent,
is an element of the set ${\cal M}/\Omega$, and vice versa,
if $(I(x,u);J(x,u))$~is a representative of an equivalence class from
${\cal M}/\Omega$, then the system $QI^q=0$, $q=\overline{1,m+n-1}$, $QJ=1$
is a determined system of linear algebraic equations with respect to the coefficients of the
corresponding operator $Q$.}

\section{The case {\mathversion{bold}$n=1$}}
Let us consider in more detail the case of one independent
variable $x$ ($n=1$), for which it is possible to obtain a
more compact formulation of Theorem~1 and of its corollaries,
and also to obtain some additional results.

\medskip

\noindent
{\bf Theorem {\mathversion{bold}$1'$}.} {\it Let $I=I(x,u)=(I^1(x,u),I^2(x,u), \ldots, I^m(x,u))$ be a
universal invariant of the operator $Q$ and
$J(x,u)$ be a particular solution of the equation $QJ=1$. Then the function
\[
I^j(x,u), \qquad \left(\frac{1}{D_xJ}D_x\right)^sI^j(x,u), \quad s=\overline{1,r},
\]
where $D_x=\p_x+u^i_x\p_{u^i}+u^i_{xx}\p_{u^i_x}+\cdots$~is the operator
of total derivative with respect to the variable~$x$,
form a complete set of $r$-th order functionally independent differential
invariants (or a universal differential invariant)
of the operator $Q$.}

\medskip

\noindent
{\bf Corollary {\mathversion{bold}$1'$}.} {\it For any operator $Q$ there exists
a complete set of $n$-th order functionally independent differential invariants,
where every invariant is a rational function of the variables
$u^i_x$, $u^i_{xx}$, \dots, $(u^i)^{(n)}$ of the jet space
with coefficients depending on $x$ and $u^i$.}

\medskip

\noindent
{\bf Corollary {\mathversion{bold}$2'$}.} {\it If $I=(I^1(x,u),I^2(x,u),
\ldots, I^m(x,u))$~is a universal invariant of the operator $Q$  and
$J=J(x,u)$~is a particular solution of the equation $QJ=1$, then the functions
\be{diff.invariant.first.order}
I^j_{(1)}=I^j_{(1)}(x,u_{(1)})=\frac{dI^j}{dJ^{\phantom j}}=\frac{D_xI^j}
{D_xJ^{\phantom j}}
=\frac{I^j_x+I^j_{u^i}u^i_x}{J_x+J_{u^{i'}}u^{i'}_x}
\ee
form a complete set of functionally independent
differential invariants of exactly first order for the operator $Q$.}

\medskip

\noindent
{\bf Corollary 3.} {\it The components of universal
differential invariants having exactly order one
of the operator $Q$ may be sought for in the form
of fractional-linear functions of the variables $u^i_x$ of the jet space
with coefficients depending on $x$ and $u^i$.}

\medskip

Corollary $2'$ may be restated using the notion
of the invariant differential.

\medskip

\noindent
{\bf Corollary 4.} {\it The ratio of invariant differentials of the operator
$Q$ of first and second type is its differential invariant of exactly
first order. If $dI^1$, $dI^2$, \dots, $dI^m$
form a complete set of independent invariant differentials of the first type
for the operator $Q$, then its ratio with its invariant differential of
the second type exhaust functionally independent differential
invariants of exactly first order for the operator $Q$.}

\medskip

\noindent
{\bf Corollary 5 (Lie Theorem) \cite{lie,ibragimov,olver.eng}.}
{\it Let $n\!=\!m\!=\!1$,
$I(x,u)$  and~$I_{(1)}(x,u,u_x)$~are differential invariants
of zero and of exactly first order for the operator $Q$.
Then the functions
\[
I, \quad I_{(1)}, \quad \frac{d^s I_{(1)}}{dI^s}=\left(\frac{1}{D_x I}D_x\right)^s I_{(1)},
 \quad
s=\overline{1,r-1},
\]
form a complete set of $n$-th order functionally independent differential invariants
 for the operator $Q$. }

\medskip

The operators of $G$-invariant differentiation for the case of one independent
variable are traditionally sought for in the form
\[
{\cal D}=\frac{1}{D_x I^0}D_x,
\]
where $I^0$~is a  differential invariant for the group $G$ (see e.g. Corollary~5).
Therefore, for the construction of an arbitrary-order universal differential invariant
for a one-parameter group of local transformations by means of the operator of
$G$-invariant differentiation of this form it is necessary to know
$m+1$ functionally independent differential  invariants for the group  $G$
of the possibly minimal, or $m$ functionally independent
zero-order differential invariants (or simply invariants), and
one exactly first-order differential invariant.
The algorithm suggested in the Theorem~$1'$ allows to avoid direct construction of differential invariants.

\medskip

\noindent
{\bf Example 1.} (Cf. \cite{olverEI&S95,olver.eng}.)
Let $n=m=1$ and $G={\rm SO}(2)$ be a group of rotations acting on $X\times U\simeq \R^2$,
with an infinitesimal operator  $Q=u\p_x-x\p_u$.
$I=\sqrt{x^2+u^2}$ is an invariant of the group $G$ (of the operator $Q$),
 whence (in notation from the proof of Theorem~2) $U(x,C)=\pm\sqrt{C^2-x^2}$. Then
\[
J=\pm\int \frac{dx}{\sqrt{C^2-x^2}}=\pm\arcsin \frac{x}{C}=\pm\arcsin \frac{x}{\sqrt{x^2+u^2}}
\]
(here we put the integration constant to be zero), whence
\[
I_{(1)}=\frac{I_x+I_uu_x}{J_x+J_uu_x}=\frac{x+uu_x}{-u+xu_x}\,\sqrt{x^2+u^2},
\quad\! \mbox{or} \quad\! \widetilde I_{(1)}=\frac{x+uu_x}{-u+xu_x}
\]
is a first-order differential invariant for the operator $Q$.

\section{The Standard Approach\\
and Integration of Riccati-Type Systems}
Within the framework of direct method the differential invariants
having exactly first order are found as invariants of the first prolongation
\[
Q^{(1)}=\xi^a\p_{x_a}+\eta^i\p_{u^i}+
(\eta^k_c+\eta^k_{u^j}u^j_c-\xi^b_{x_c}u^k_b-\xi^b_{u^j}u^j_cu^k_b)\p_{u^k_c}
\]
of the operator $Q$, or as the first integrals of the corresponding
characteristic system of ordinary differential equations
\be{characteristic.system}
\frac{dx_a}{\displaystyle \xi^a\vphantom{\displaystyle \eta^i}}=\frac{du^i}{\displaystyle \eta^i}=
\frac{du^k_c}{\displaystyle \eta^k_c+\eta^k_{u^j}u^j_c-\xi^b_{x_c}u^k_b-\xi^b_{u^j}u^j_cu^k_b},
\ee
that depend not only on $x$ and $u$, but also on the other the variables of the space $J_{(1)}$.
(Here $u^i_a$~is a variable of the jet space $J_{(1)}$,
which corresponds to the derivative
$\p u^i/\p x_a$; lower indices of functions stand for the derivatives with respect
to the corresponding variables;
there is no summation over  $a$, $c$, $i$ and $k$ in the latter equation).
Integration of the system (\ref{characteristic.system}) is, as a rule, a highly cumbersome task.
If a universal invariant~$I(x,u)$ for the operator $Q$ is known, then it amounts
to integration of Riccati-type systems of the form
\be{system.of.Riccati.equations.for.first.diff.invariant.a}
\frac{du^k_c}{dx_a}=-\frac{\xi^b_{u^j}}{\xi^a}u^j_cu^k_b+\frac{\eta^k_{u^j}}{\xi^a}u^j_c
-\frac{\xi^b_{x_c}}{\xi^a}u^k_b+\frac{\eta^k_{x_c}}{\xi^a}
\left.\phantom{\!\!\!\!\!\!\frac{C}{C}}\right|_{\!\!\!\raisebox{1ex}[0ex][0ex]{
$\begin{array}{l}\scriptstyle u=U(x_a,C) \\[-0.7ex]\scriptstyle x_d=X^d(x_a,C),\; d\not=a \end{array}$}\!\!},
\ee
if $\xi^a\not=0$ for some fixed $a$, or
\be{system.of.Riccati.equations.for.first.diff.invariant.b}
\frac{du^k_c}{du^i}=-\frac{\xi^b_{u^j}}{\eta^i}u^j_cu^k_b+\frac{\eta^k_{u^j}}{\eta^i}u^j_c
-\frac{\xi^b_{x_c}}{\eta^i}u^k_b+\frac{\eta^k_{x_c}}{\eta^i}
\left.\phantom{\!\!\!\!\!\!\frac{C}{C}}\right|_{\!\!\!\raisebox{1ex}[0ex][0ex]{
$\begin{array}{l}\scriptstyle x=X(u^i\!,C),\\[-0.7ex]\scriptstyle u^l=U^l(u^i\!,C),\; l\not=i\end{array}$}\!\!},
\ee
if $\eta^i\not=0$ for some fixed $i$. Here $x_d=X^d(x_a,C)$, $d\not=a$, $u=U(x,C)$
and $x=X(u^i,C)$, $u^l=U^l(u^i,C)$, $l\not=i$,~are solutions of the system of algebraic equations
$I(x,u)=C$ with respect to the variables $x_d$, $d\not=a$, $u$ and $x$, $u^l$, $l\not=i$,
respectively.
The constants $C=(C_1,C_2,\ldots,C_{m+n-1})$ in the
systems~(\ref{system.of.Riccati.equations.for.first.diff.invariant.a})
and~(\ref{system.of.Riccati.equations.for.first.diff.invariant.b})
are considered as parameters.
The case $\eta^i\not=0$ could be reduced to the case $\xi^a\not=0$ by means of
the locus transformation:
\bss
\tilde x_a=u^i, \quad \tilde x_d=x_d, \quad
\tilde u^i=x_a, \quad \tilde u^l=u^l,\quad   d\not=a,\quad l\not=i,\\[1ex] \displaystyle
\tilde u^i_a=\frac{1}{u^i_a}, \quad \tilde u^i_d=-\frac{u^i_d}{u^i_a}, \quad
\tilde u^l_a=\frac{u^l_a}{u^i_a}, \quad \tilde u^l_d=u^l_d-\frac{u^i_d}{u^i_a}u^l_a. \quad
\ess
For this reason we will consider in detail only the case $\xi^a\not=0$.

The method we suggested in Corollary 2 for finding differential
invariants having exactly first order, unlike the standard method, allows to avoid
direct integration of systems of Riccati
equations~(\ref{system.of.Riccati.equations.for.first.diff.invariant.a})
or~(\ref{system.of.Riccati.equations.for.first.diff.invariant.b})
and to find a solution through one quadrature and differentiation.
This result means that {\it in the case of known universal
invariant~$I(x,u)$ for the operator $Q$ the systems~(\ref{system.of.Riccati.equations.for.first.diff.invariant.a})
and~(\ref{system.of.Riccati.equations.for.first.diff.invariant.b}) are always integrable by means of one quadrature}.
Really, the general solution for the system~(\ref{system.of.Riccati.equations.for.first.diff.invariant.a})
could be given explicitly by $m$ non-linked systems of linear algebraic equations
\[
D_{x_b}\hat I^j
\left.\vphantom{\frac{C}{C}}\right|_{\!\raisebox{1ex}[0ex][0ex]{%
$\begin{array}{l}\scriptstyle u=U(x_a,C) \\[-0.7ex]\scriptstyle x_d=X^d(x_a,C),\; d\not=a \end{array}$}\!\!}  =0,
\]
where $\hat I^j=I^{j+n-1}+\sum_{d=1}^{n-1}\widetilde C_{jd}I^d+\widetilde C_{jn}J$,
$\widetilde C_{ib}$~are arbitrary constants.
To write the solution in the explicit form, we introduce some additional
notation:
\bss
\bar x=(x_d)_{d=1,\,d\not=a}^n, \quad \bar X=(X^d)_{d=1,\,d\not=a}^n, \quad z=x_a,\\[1.5ex]
I^{\bar x}=(I^d)_{d=1}^{n-1}, \quad I^{u}=(I^{j+n-1})_{j=1}^m,  \\[1.5ex]
C^{\bar x}=(C_d)_{d=1}^{n-1}, \quad C^{u}=(C_{j+n-1})_{j=1}^m,  \\[1.5ex]
\widetilde C'=(\widetilde C_{jd})_{j=1\vphantom{d}}^m\,{}_{d=1\vphantom{j}}^{n-1}, \quad
\widetilde C''=(\widetilde C_{jn})_{j=1}^m, \quad \hat I=I^u+\widetilde C'I^{\bar x}+\widetilde C''J.
\ess
Then the general solution for the system~(\ref{system.of.Riccati.equations.for.first.diff.invariant.a})
is given by the formulae
\[
(u^j_b)_{j=1}^m\,{}_{b=1}^n=-\hat I_u^{-1}\hat I_x
\left.\vphantom{\frac{C}{C}}\right|_{\!\raisebox{1ex}[0ex][0ex]{%
$\begin{array}{l}\scriptstyle u=U(x_a,C) \\[-0.7ex]\scriptstyle \bar x=\bar X(x_a,C) \end{array}$}\!\!},
\]
or
\bss
(u^j_a)_{j=1}^m=U_z-U_{C^{\bar x}}\bar X_{C^{\bar x}}^{-1}\bar X_z
+H ( (\widetilde C'+\widetilde C''J_{C^{\bar x}})\bar X_{C^{\bar x}}^{-1}\bar X_z-\widetilde C''J_{C^{\bar x}} ),
\\[1.5ex]
(u^j_b)_{j=1\vphantom{b}}^m\,{}_{b=1,\,b\not=a\vphantom{j}}^n=
U_{C^{\bar x}}\bar X_{C^{\bar x}}^{-1}-H(\widetilde C'+\widetilde C''J_{C^{\bar x}})\bar X_{C^{\bar x}}^{-1},
\ess
where $H=(U_{C^u}-U_{C^{\bar x}}\bar X_{C^{\bar x}}^{-1}\bar X_{C^u})
(E+\widetilde C''J_{C^u}-(\widetilde C'+\widetilde C''J_{C^{\bar x}})\bar X_{C^{\bar x}}^{-1}X_{C^u})^{-1}$,
$E$~is the  $m\times m$ unit matrix;
the signs of vector-functions with lower indices of the sets of variables designate the
corresponding Jacobi matrices.
To ensure the existence in some neighborhood of a fixed point $(x^0,u^0)$
of all inverse matrices,
that are mentioned above, it is sufficient to consider constants
 $\widetilde C_{ib}$ to be small and
to perform a (preliminarily determined as non-degenerate) linear change
in the set of invariants for the matrix  $I_{(\bar x,u)}(x^0,u^0)$
to be the unit matrix.

If we put $\widetilde C_{ib}=0$, then we obtain a particular solution
\[
(u^j_a)_{j=1}^m=U_z-U_{C^{\bar x}}\bar X_{C^{\bar x}}^{-1}\bar X_z, \quad
(u^j_b)_{j=1\vphantom{d}}^m\,{}_{b=1,\,b\not=a\vphantom{j}}^n=U_{C^{\bar x}}\bar X_{C^{\bar x}}^{-1}.
\]

The solution of the system~(\ref{system.of.Riccati.equations.for.first.diff.invariant.a})
in explicit form should be written separately for the case
$n=1$. As in this case $u=U(x,C)$~is the general solution of the
system~$du^j/dx=\eta^j(x,u)/\xi(x,u)$, then it is easy to verify that $u_x=U_x(x,C)$
is a particular solution for the system~(\ref{system.of.Riccati.equations.for.first.diff.invariant.a})
(here, as in~(\ref{system.of.Riccati.equations.for.first.diff.invariant.a}), $C$~is a set of parameters).
The general solution for the system~(\ref{system.of.Riccati.equations.for.first.diff.invariant.a})
has the form
\be{ux.m.equal.1}
u_x=-(I_u-\widetilde C \otimes J_u)^{-1}(I_x+\widetilde CJ_x) \biggl|_{u=U(x,C)}=
U_z-U_C(E+\widetilde C \otimes J_C)^{-1}\widetilde CJ_z
\ee
(in the latter equality the substitution  $x=z$, $u=U(z,C)$ was performed), where
$E$~is the unit matrix of the dimensions~$m\times m$,
$\widetilde C=(\widetilde C_1,\widetilde C_2,\ldots,\widetilde C_m)^{\rm T}$~is a column of arbitrary constants,
$I_u=(I^i_{u^j})$, $I_x=(I^i_x)$, $U_z=(U^k_z)$, $U_C=(U^i_{C_j})$,
$\widetilde C\otimes J_u=(\widetilde C^kJ_{u^l})$, 
$\widetilde C\otimes J_C=(\widetilde C^kJ_{C^l})$. 
The inverse matrices in~(\ref{ux.m.equal.1}) always exist for sufficiently small
$\widetilde C_i$.

\medskip

\noindent
{\bf Example 2.} Let $n=m=1$, $Q=\exp(-x-u)(\partial_x+u\partial_u)$.
$I(x,u)=u\exp(-x)$ is an invariant for the operator $Q$,
whence $U(x,C)= C\exp(x)$. Then
\[
J=\int \frac{dx}{\exp(-x-C\exp(x))}=
\frac 1C \exp(C\exp(x)) =\frac{\exp(x+u)}{u}
\]
(here we set the integration constant to be zero). Therefore,
\[
I_{(1)}=\frac{I_x+I_uu_x}{J_x+J_uu_x}=\frac{\exp(-2x-u) u^2 (u_x-u)}
{u+uu_x -u_x} \quad \mbox{or}
 \quad \widetilde I_{(1)}=\frac{u_x-u}{u+uu_x-u_x}\exp(-u)
\]
is a first-order differential invariant for the operator $Q$.
System~(\ref{system.of.Riccati.equations.for.first.diff.invariant.a}) for the operator $Q$
consists of one equation which has the form
\[
\frac{d u_x}{dx}=u_x^2+(2-C \exp(x))u_x- C\exp(x).
\]
The function $v=C \exp(x)$ is its particular solution.
The general solution
of this Riccati equation is given by the formula
\[
u_x=C\exp(x) -\frac{C^2\exp(2x)}{C\exp(x)-1+\widehat C \exp(-C\exp(x))},
\]
where  $\widehat C$ is an arbitrary constant.

\vspace{1ex}

\noindent
{\bf Example 3.} Let $n=m=1$, $Q=xu(x\partial_x+ k u\partial_u)$, $k\in\R$.
$I(x,u)=u x^{-k}$ is an invariant for the operator  $Q$,
whence $U(x,C)= C x^k$. Then
\[
\displaystyle
J=\!\int\!\!\! \frac{dx}{C x^{k+2}}=\left\{\!\begin{array}{ll}
\displaystyle \frac{\ln x}{C}= \frac{\ln x}{xu}, & \mbox{if} \quad k=-1,
\\[2.5ex]
\displaystyle -\frac{x^{-(k+1)}}{(k+1)C}= -\frac{1}{(k+1)x u}, &
\mbox{if} \quad k\not=-1
\end{array}\right.
\]
(here we set the integration constant to be zero).
The corresponding Riccati equation has the form
\[
\frac{d u_x}{dx} =-\frac{1}{Cx^k} u_x^2+\frac{2(k-1)}{x} u_x+kCx^{k-2}.
\]
The function $u_x=kCx^{k-1}$ is its particular solution. The general solution
of this Riccati equation is given by the formulae
\[
u_x=-\frac{C}{x^2} \left(1+\frac{1}{\widehat C -\ln x}\right),
\quad \mbox{if}\;  k=-1, \quad \mbox{or}
\]
\[
u_x=Cx^{k-1}\left(k-\frac{k+1}{1+\widehat C x^{k+1}}\right),
\quad \mbox{if}\; k\not=-1,
\]
where $\widehat C$ is an arbitrary constant.

\vspace{1ex}

\noindent
{\bf Remark.} For well-known transformation groups on
the plane (i.e. $n=m=1$) integrability in quadratures of
equations~(\ref{system.of.Riccati.equations.for.first.diff.invariant.a})
and~(\ref{system.of.Riccati.equations.for.first.diff.invariant.b})
as a rule obviously follows from the form of these equations. For
instance, when $\xi_u=0$ or $\eta_x=0$,
they are a linear equation or a Bernoulli equation respectively.
If $G$ is a one-parameter group of conformal transformations, then
$\xi_x=\eta_u$ and $\xi_u=-\eta_x$, and therefore
in equations~(\ref{system.of.Riccati.equations.for.first.diff.invariant.a})
and~(\ref{system.of.Riccati.equations.for.first.diff.invariant.b}) variables
are separated:
\[
\frac{dv}{v^2+1}=\frac{\eta_x}{\xi}dx\left.\phantom{\!\!\!\!\!\frac{C}{C}}\right|_{u=U(x,C)}
\qquad\mbox{and}\qquad
\frac{dv}{v^2+1}=\frac{\eta_x}{\eta}du\left.\phantom{\!\!\!\!\!\frac{C}{C}}\right|_{x=X(u,C)}.
\]

\vspace{1ex}

\noindent
{\bf Example 4.} Let $n=1$, $m=2$, $Q=\exp(-x-u^1-u^2)(\partial_x+
u^1\partial_{u^1}+u^2\partial_{u^2})$.
$I^1(x,u^1,u^2)=u^1 \exp (-x)$ and
$I^2(x,u^1,u^2)=u^2 \exp (-x)$
are invariants for the operator  $Q$,
whence $U^1(x,C^1,C^2)= C^1 \exp(x)$ and $U^1(x,C^1,C^2)= C^2 \exp(x)$. Then
\[
\displaystyle
J\left(x,C^1,C^2\right)=\!\int\!\!\! \frac{dx}{\exp\left(-x-\left(C^1+C^2\right)\exp(x)\right)}=
\frac{\exp\left(\left(C^1+C^2\right)\exp(x)\right)}{C^1+C^2}.
\]
(here we set the integration constant to be zero).
The corresponding Riccati-type system has the form
\[
\begin{array}{l}
\displaystyle \frac{d u^1_x}{dx} =\left(u_x^1+u_x^2\right)u_x^1 +
\left(2-C^1 \exp(x)\right) u_x^1 -C^1 \exp(x) u_x^2 -C^1 \exp(x),
\vspace{2mm}\\
\displaystyle \frac{d u^2_x}{dx} =\left(u_x^1+u_x^2\right)u_x^2  -C^2 \exp(x) u_x^1
+\left(2-C^2 \exp(x)\right) u_x^2 -C^2 \exp(x).
\end{array}
\]
According to (\ref{ux.m.equal.1}), the general solution
of this system is given by the formula
\[
\begin{array}{l}
\displaystyle
\left(\begin{array}{c} u_x^1 \vspace{2mm}\\ u_x^2\end{array}\right)=
\exp(x)
\left(\begin{array}{c} C^1 \\ C^2\end{array}\right)+
\vspace{2mm}\\
\displaystyle \qquad \qquad {}+
\frac{\left(C^1+C^2\right)\exp(2x) J\left(x,C^1,C^2\right)}
{1-\left(\widetilde C^1+\widetilde C^2\right)\left(\exp(x) -\left(C^1+C^2\right)^{-1}\right)
J\left(x,C^1,C^2\right)}
\left(\begin{array}{c} \widetilde C^1 \\ \widetilde C^2\end{array}\right),
\end{array}
\]
where $\widetilde C^1$, $\widetilde C^2$ are arbitrary constants.

\vspace{1ex}

\noindent
{\bf Example 5.} Let $n=1$, $m=2$, $Q=\exp(u^1+u^2)(\partial_x+
u^2\partial_{u^1}-u^1\partial_{u^2})$.
Then, 
\[
\begin{array}{l}
\displaystyle U^1\left(x,C^1,C^2\right)= C^1 \cos x +C^2 \sin x,
\vspace{2mm}\\
\displaystyle U^2\left(x,C^1,C^2\right)= -C^1 \sin x +C^2 \cos x,
\vspace{2mm}\\
\displaystyle J\left(x,C^1,C^2\right)=\int\exp\left(-\left(C^1+C^2\right)\cos x-\left(C^2-C^1\right)\sin x
\right) dx.
\end{array}
\]
The corresponding Riccati-type system has the form
\[
\begin{array}{l}
\displaystyle \frac{d u^1_x}{dx} =-\left(u_x^1+u_x^2\right)u_x^1 +
\left(-C^1\sin x +C^2\cos x\right) u_x^1 + \left(-C^1\sin x +C^2\cos x+1\right) u_x^2,
\vspace{2mm}\\
\displaystyle \frac{d u^2_x}{dx} =-\left(u_x^1+u_x^2\right)u_x^2  -
\left(C^1\cos x +C^2\sin x+1\right) u_x^1 - \left(C^1\cos x +C^2\sin x\right) u_x^2.
\end{array}
\]
It follows from (\ref{ux.m.equal.1}) that the general solution
of this system is given by the formula
\[
\begin{array}{l}
\displaystyle
\left(\begin{array}{c} u_x^1 \vspace{2mm}\\ u_x^2\end{array}\right)=
\left(\begin{array}{c} -C^1\sin x +C^2 \cos x \\ -C^1\cos x-C^2\sin x \end{array}\right)+
\vspace{2mm}\\
\displaystyle \qquad \qquad {}+ \frac{\exp\left(-\left(C^1+C^2\right)\cos x-\left(C^2-C^1\right)\sin x
\right)}{1-\widetilde C^1 J_{C^1}-\widetilde C^2 J_{C^2}}\left(\begin{array}{cc}
\widetilde C^1\cos x +\widetilde C^2 \sin x \\
-\widetilde C^1\sin x +\widetilde C^2 \cos x\end{array}\right),
\end{array}
\]
where $\widetilde C^1$, $\widetilde C^2$ are an arbitrary constant.

\subsection*{Acknowledgments}
The authors thank Dr. I.A. Yehorchenko and Dr. A.G. Sergyeyev for the fruitful discussion of the
results of this paper.

\end{document}